# Adaptive Subspace Sampling for Class Imbalance Processing-Some clarifications, algorithm, and further investigation including applications to Brain Computer Interface

Chin-Teng Lin, *Fellow, IEEE*, Kuan-Chih Huang, Yu-Ting Liu, Yang-Yin Lin, Tsung-Yu Hsieh
Nikhil R. Pal, *Fellow, IEEE*, Shang-Lin Wu, Chieh-Ning Fang, Zehong Cao

Chin-Teng Lin
*Center for Artificial Intelligence and Faculty of Engineering and Information Technology*
*University of Technology Sydney*
Sydney, Australia
Chin-Teng.Lin@uts.edu.au

Kuan-Chih Huang
*Department of Electrical and Computer Engineering and Brain Research Center*
*National Chiao Tung University*
Hsinchu, Taiwan
kchuang.ece91g@nctu.edu.tw

Nikhil R. Pal
*Electronics and Communication Sciences Unit*
*Indian Statistical Institute*
Calcutta, India
nikhil@isical.ac.in

Zehong Cao, Yu-Ting Liu
Chieh-Ning Fang
*Center for Artificial Intelligence and Faculty of Engineering and Information Technology*
*University of Technology Sydney*
Sydney, Australia

Tsung-Yu Hsieh
*College of Information Sciences and Technology*
*University of Pennsylvania*
Philadelphia, USA

Yang-Yin Lin, Shang-Lin Wu
*Brain Research Center*
*National Chiao Tung University*
Hsinchu, Taiwan

*Abstract*—Kohonen's Adaptive Subspace Self-Organizing Map (ASSOM) learns several subspaces of the data where each subspace represents some invariant characteristics of the data. To deal with the imbalance classification problem, earlier we have proposed a method for oversampling the minority class using Kohonen's ASSOM. This investigation extends that study, clarifies some issues related to our earlier work, provides the algorithm for generation of the oversamples, applies the method on several benchmark data sets, and makes an application to a Brain Computer Interface (BCI) problem. First we compare the performance of our method using some benchmark data sets with several state-of-the-art methods. Finally, we apply the ASSOM-based technique to analyze a BCI based application using electroencephalogram (EEG) datasets. Our results demonstrate the effectiveness of the ASSOM-based method in dealing with imbalance classification problem.

## I. INTRODUCTION

Learning from imbalanced data has attracted growing attention in the research community in recent years because it arises in many application problems, including medical diagnosis, anomaly detection, and financial fraud detection [1-4]. Although many attempts have been made to tackle this problem, there are still challenges to be addressed. Specifically, a classification task can be regarded as an imbalanced problem whenever the number of samples in one or more classes significantly differ from those of the other classes.

In this paper, for simplicity, we focus on the two-class imbalanced classification problem, which is a topic of major interest in the research community. The imbalance can be viewed in two common forms: relative imbalance and absolute imbalance. Relative imbalance occurs when minority samples are well represented but severely outnumbered by the majority of samples, whereas absolute imbalance arises in datasets in which minority samples are scarce and underrepresented. Either form of imbalance poses a great challenge to conventional classification algorithms because it becomes extremely difficult to detect minority class samples. Since most classifiers minimize the misclassification error in some form or other, in an imbalanced case the classifiers tend to favor the majority class and sometimes even effectively omit the minority class samples in the training process, and thereby results in a biased classifier. This causes severe problem when the detection of minority class samples is crucially important, such as in cancer diagnosis.

Current solutions to the imbalanced problem can be divided into two categories: internal methods and external methods. Internal methods target the imbalanced problem by modifying the underlying classification algorithm. A popular approach in this category is cost-sensitive learning [5], which uses a cost matrix for different types of errors or instances to facilitate the learning directly from an imbalanced dataset. A higher cost of misclassifying a minority class sample compensates for the scarcity of the minority class. In [6], a cost-sensitive framework for applying the support vector machine is proposed. In [7], Zhou and Liu investigated the applicability of cost-sensitive neural networks on the imbalanced

This study is funded by the Ministry of Science and Technology of the Republic of China, Taiwan, under contract no. MOST 108-2221-E-009-120 -MY2.

classification problem. In contrast, external methods aim to address the imbalanced problem by manipulating the input data to form a more balanced data set. External methods can further be divided into under-sampling and oversampling. Under-sampling methods compensate for the imbalanced problem by reducing the instances of the majority class. cluster-based under-sampling approach is proposed in [8]. There are studies that demonstrate that class cover catch diagrams capture the density of majority class as radii of the covering balls preserve the information during the under-sampling process. In contrast to under-sampling methods that remove majority class samples, oversampling methods balance the data set by generating synthetic samples for the minority classes. The synthetic minority oversampling technique (SMOTE) [9] algorithm generates synthetic minority samples to eliminate the classifier learning bias. Several extension of SMOTE algorithm has been proposed, e.g., the Borderline-SMOTE[10], SMOTE-Boost[11], majority weighted minority oversampling technique (MWMOTE)[12] and adaptive synthetic sampling (ADASYN)[13]. In [14], an enhanced structure-preserving oversampling (ESPO) method that is based on a combination of the multivariate Gaussian distribution and an interpolation-based algorithm is developed.

Kohonen in [15, 16, 30] proposed a special type of Self-Organizing Map called the Adaptive Subspace Self-Organizing Map (ASSOM). The ASSOM consists of different modules where each module learns to recognize invariant patterns that are subjected to simple transformation (each module represents a subspace). Thus, each subspace represents some invariant characteristics of a subset of the data. It is, therefore, reasonable to assume that if we can generate synthetic instances from each of these subspaces, these instances will follow the distribution of the original data. The ASSOM concept has been extended to propose Kernel Adaptive Subspace Self-organizing map (KASSOM)[17, 18]. In [19], we have used the KASSOM concept to develop an effective algorithm for minority oversampling. Unlike the method in [19] which exploits the properties of KASSOM, in [20] we have exploited directly the properties of Kohonen's ASSOM in the data space (not in the kernel space) to deal with imbalanced classification problem. In [20] our contribution was simply the use of Kohonen's ASSOM to develop an algorithm for sampling the minority classes. Due to lack of proper referencing this was not clear in [20]. This paper clarifies this point, extends the study in by providing a clear implementable algorithm for generation of the samples, making a more detailed investigation with more data sets. Typically, the objective function uses the usual kernel function considering the position of the winner and non-winner modules on the ASSOM layer, but here we use a different kernel function that is consistent with our objectives. More importantly, in this work we demonstrate the effectiveness of the ASSOM based algorithm on an important imbalanced Brain Computer Interface (BCI) problem using electroencephalogram (EEG) data.

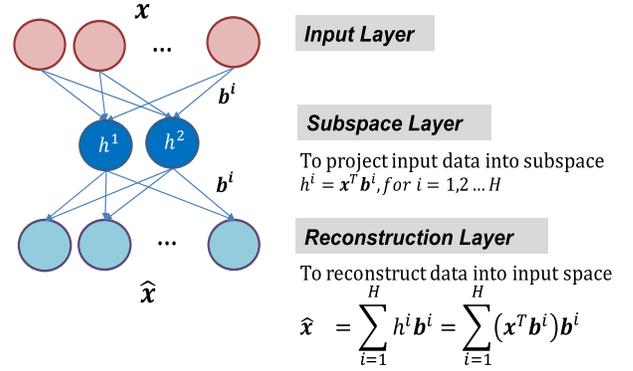

Fig. 1. The architecture of the ASSOM

## II. KOHONEN'S ADAPTIVE-SUBSPACE SELF-ORGANIZING MAP (ASSOM)

First we shall briefly discuss ASSOM [15,16,30] and then we shall explain the sampling algorithm based on ASSOM. Fig. 1 shows the architecture of ASSOM. This is a special type of self-organizing map. Conventional SOM finds prototypes which are representative of the training data, typically each prototype represents a group of data points which are similar. Also, the prototypes, which are spatially closer on the map, are similar. In case of ASSOM, instead of prototypes, it can find translation, rotation, and scale invariant subspaces/filters. It finds subspaces, where each subspace represents some invariant characteristics of the training data. Thus one can view it as an alternative to PCA type feature extractor. In ASSOM each invariant class / group is represented by a two-layer neural architecture or module. In Fig. 1, the second layer nodes are the modules, they are called quadratic neurons /modules as we shall see that they minimizes a quadratic error. And each module is responsible for an invariant subspace. The third layer in Fig. 1 is the reconstruction layer. Fig. 1 gives the overall idea of the architecture which is further detailed in Fig. 2. Let $x$ be an input vector (signal). Given a set of input vectors, ASSOM finds linear subspaces of dimension H so that the original signal can be reconstructed from the projections. A linear subspace $L$ of dimensionality $H$ is defined given a set of linearly independent basis vectors $b_1, \ldots b_{l_i}$, and the reconstructed signal is obtained using Eq. (1). Like ordinary SOM, ASSOM learning algorithm also uses gradient based learning for estimation of the basis vectors.

Each node in layer 2 can be viewed as a representing linear-subspace neural unit. Each node represents a linearly independent basis vector. The output function of layer 3, which reconstructs the input is written as

$$\hat{x} = \sum_{i=1}^{H} x b_i b_i^T \qquad (1)$$

where $x$ denotes input data, $b_i$ denotes a basis vector in the orthonormal form and $H$ denotes the number of hidden nodes. For ortho-normalization, the Gram-Schmidt process is used. The reconstructed signal relies on the orthonormal basis; in other words, the reconstructed signal $\hat{x}$ that belongs to $L$ is the orthogonal projection of $x$ onto $L$.

We expect that the reconstructed signal is approximately similar to the original signal; thus, the network tries to minimize the reconstruction error $\|\widetilde{x}\| = \|x-\hat{x}\|^2$. Finally, the projection operator matrix $P$ can be defined as in Eq. (2), and the following properties hold: $P^2 = P$ and $P^T = P$.

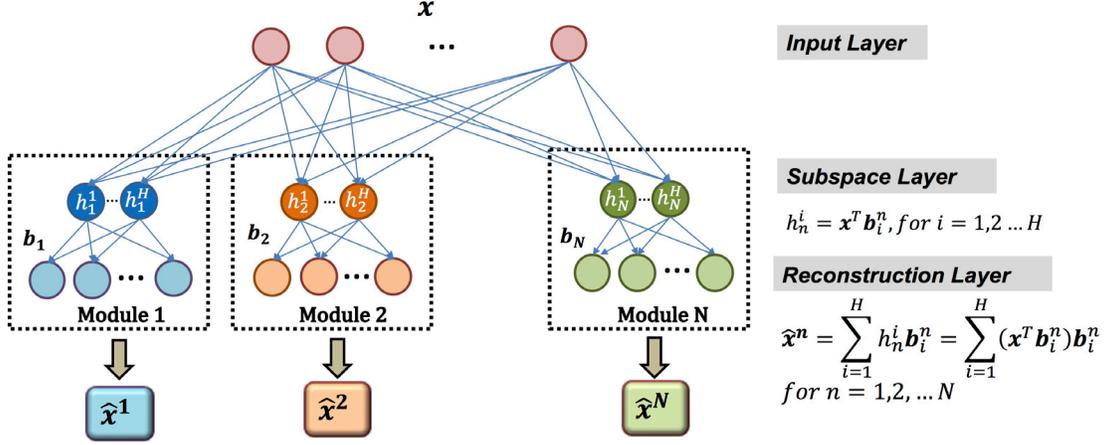

Fig. 2. Competitive learning of ASSOM

$$P = \sum_{i=1}^{H} b_i b_i^T \quad (2)$$

where

$$\hat{x} = Px \text{ and } \widetilde{x} = (I-P)x, \quad (3)$$

in which $I$ represent the identity matrix.

### A. Learning Scheme

The mechanism of the ASSOM is inherited from SOM and it uses a similar competitive learning for the basis vectors, which are vital component to the effectiveness and robustness of the system. As shown in Fig. 2, each module computes the projected outputs and then the square error of reconstruction. This error is the basis for the competition. The module which best reconstructs the input will be the winner module and will update its basis vectors. Like SOM it will also update the basis vectors of its spatially neighboring modules. The details of the update produce are described more or less in line with [15, 16, 30].

The competitive learning is described in Fig. 2. The different modules compete on the input signal to find the minimum distance module as the winner, which corresponds to the best subspace whose projection can reconstruct the input in the best way in terms of square error.

Thus the winner module is defined by:

$$c = argmin_{n \in N} \left\{ \sum_{t \in S} \|x - \widetilde{x^n(t)}\|^2 \right\}$$

$$= argmin_{n \in N} \left\{ \sum_{t \in S} \|\widetilde{x^n(t)}\|^2 \right\} \quad (4)$$

where $S$ is denoted as the total number of input samples and $c$ as the index of the winning module. In order to define the error function, authors in [15, 16] proposed a multiplier for the error. In [30] to modulate the strength of update as we move away from the winner, authors suggested to use a neighborhood kernel which decreases with the distance between the winning module and the neighboring module on the ASSOM array. Here since our objective is to have good reconstruction, we use a kernel function based on the actual distances between the reconstructed input by the winner and its neighbor, $g_c^n(t)$, as follows:

$$g_c^n(t) = exp\left[-\frac{\|\widetilde{x^c(t)} - \widetilde{x^n(t)}\|^2}{2\sigma^2}\right] \quad (5)$$

where $\sigma$ is a constant. We note that typically the kernel function suggested in [30] is used.

Cost function $E$ is defined as the summation of projection error for all modules and data.

$$E = \sum_n \sum_{t \in S} g_c^n(t) \|\widetilde{x^n(t)}\|^2 \quad (6)$$

By using the gradient descent (GD) algorithm for each input sample, the update equation for basis vectors of each module becomes

$$b_i^n(t+1) = b_i^n(t) - \eta \frac{\partial E}{\partial b_i^n(t)} \quad (7)$$

where $b_i^n$ is the $i^{th}$ basis vector of module $n$ and the factor $\eta$ is a learning rate, and the derivation is computed as

$$\frac{\partial E}{\partial b_i^n(t)} = -2 \sum_{t \in S} g_c^n(t) x(t) x(t)^T b_i^n(t) \quad (8)$$

Based on Eqs. (7) and (8), the basis vectors are updated as follows:

$$b_i^n(t+1) = [I + \eta g_c^n(t) x(t) x(t)^T] b_i^n(t) \quad (9)$$

Authors in [30] suggested that the magnitude of correction should be an increasing function of the error. In order to guarantee this authors [30] suggested to divide the learning rate by the scalar $\|\hat{x}^n\| / \|x\|$. Following the same principle, and denoting the learning rate as $\lambda$ we can rewrite the update rule Eq. (9) as

$$b_i^n(t+1) = \left[I + \bar{\lambda}^n(t) \frac{x(t)x(t)^T}{\|\hat{x}^n(t)\| / \|x(t)\|}\right] b_i^n(t) \quad (10)$$

where $\lambda^n(t) = \eta g_c^n(t)$.

Kohonen et al. [30] suggested that instability of ASSOM can be eliminated and much better filters can be generated if during the learning process, we set the magnitude of the small components of the basis vectors $b_i^n$ to zero to reduce the degrees of freedom; thus, the $b_i^n$ is forced to approximate the dominant frequency components by a dissipation effect $\overline{b_i^n}$ which can be described by

$$\overline{b_i^n} = sgn(b_i^n) \cdot \max[0, abs(b_i^n) - \varepsilon] \quad (11)$$

where $\varepsilon$ is a small fraction of the magnitude vector that can be modeled as the following equation:

$$\varepsilon = \alpha \cdot abs[b_i^n(t) - b_i^n(t-1)] \quad (12)$$

where $\alpha$ is a small constant. The $\varepsilon$ needs to be applied after the GD algorithm is performed and prior to normalization.

To sum up, since the set of basis vectors associated with a module is required twice, once to compute Equation (1), to make an efficient network implementation, a quadratic neuron representing a module in the ASSOM has been expanded to have an additional layer (Fig. 2). Fig. 2 to has another layer of neurons so that a copy of the basis vectors is available for computation of Equation (1).

After the training of the ASSOM is over we are ready to generate the synthetic samples as detailed in the algorithm below:

**Algorithm for generation synthetic samples**

1. Suppose the ASSOM is trained with N quadratic modules.
2. Suppose we want to generate a synthetic sample from class k
3. Select a training data point **x** from class k. Get its subspace representations. There will be N such representations.
4. Inversely transform (reconstruct) the synthetic data back to the original space. So, there will be N synthetic instances. (if we want to select a few of the N, we may do so based on the reconstruction error)

Now the question comes what value of N should be chosen. One simple way may be to use the formula

$$N = round\left(\frac{\# \text{ of majority class}}{\# \text{ of minority class}}\right) - 1$$

$$= round(imbalance\ ratio) - 1$$

where *N* denotes the number of competing modules.

III. EXPERIMENTS AND RESULTS

To effectively demonstrate the performance from using different oversampling methods, we apply two popular classifiers, Multilayer Perceptron (MLP)[21] and SVMs[22]. MLPs and SVMs both play important roles in solving classification problems. We use only these two classifiers, MLP and SVM with RBF kernel (SVM-RBF), as we assume they are reliable and adequate to test the oversampling performance based on benchmark datasets and good at handling the EEG-based classification tasks.

*A. Assessment Metric*

Four commonly used assessment metrics, the recall, precision, G-mean, and F1-value, are considered to determine the benefits of the ASSOM-based algorithm for imbalanced classification problems. Four metrics are obtained by counting the number of true positive (TP), true negative (TN), false positive (FP), and false negative (FN) samples. These metrics are shown in Eqs. (13) - (16).

$$Precision = \frac{TP}{TP + FP} \quad (13)$$

$$Recall = \frac{TP}{TP + FN} \quad (14)$$

$$G\text{-}mean = \sqrt{\frac{TP}{TP + FN} \times \frac{TN}{FP + TN}} \quad (15)$$

$$F1 - value = 2\frac{Recall \times Precision}{Recall + Precision} \quad (16)$$

*B. Evaluation Results*

Eight benchmark datasets from the UCI machine learning repository [23] and KEEL datasets [24] are employed to test the ASSOM-based method and to compare with some existing relevant oversampling methods. The eight real world data sets are: Abalone, Breast cancer, E. coli, Glass, Pima, Vehicle, and Yeast. These sets are chosen such that they have different characteristics in terms of number of samples, features, classes, and imbalanced ratios. Some of these datasets have samples from more than two classes. For simplicity, these datasets are transformed into a two-class problem in this study.

There exist highly imbalanced ratios in the present two-class problems. The proposed method is evaluated by the before-and-after test to show the improvement compared to the classifiers that were constructed based on the original datasets, for which no oversampling is done. After the before-and-after test, the ASSOM is further compared to existing state-of-the-art approaches, namely, SMOTE, ADASYN, ESPO, MWMOTE, SVM-light, and SVM-balanced, to show the improvement realized by the proposed method.

For each comparative model, 70% of the data are randomly selected to use as the training set, whereas the remaining data serve as test data. To maintain the imbalanced ratio in each dataset, the selection of majority and minority samples are done maintaining original imbalance ratio of the data set.

Furthermore, for neural networks, the classification task was repeated 50 times to prevent bias in the initial state parameters during the supervised learning procedure. This overall process of validation is repeated 5 times; the reported results are the average performances.

The validation results of ANNs and SVMs with different oversampling approaches on the eight datasets are shown in Table 1 and Table 2, respectively. The best performance is shown in boldface. The results of the ANN and SVM show that the ASSOM-based method outperforms existing methods for the majority of the real-world problems.

To show the improvement of ASSOM-based method better, all of the comparative approaches are ranked based on the results of each assessment metric. Under each assessment metric, the method with the best performance is given the highest point (ANN: 6 and SVM: 8), and the worst is scored as 1 point. Consequently, we compute the average rank of the four assessment metrics across the eight datasets to quantify the relative performances. By further averaging these four assessment metrics, an overall assessment metric is used to make the comparison easier. The ANN and SVM with the best performance, which possess the highest number of points, are shown in the last row of Table 1 and Table 2. The average overall rank of the ASSOM-based method is 4.59 for the ANN and 5.66 for the SVM, which is higher than any of the other state-of-the-art approaches. These experimental results suggest that our ASSOM-based approach can yield a significant improvement in the performance of the imbalanced correction.

Table 1. Average ANN performance comparison for different comparative methods

| Dataset | Measure | Original | SMOTE | ADASYN | MWMOTE | ESPO | ASSOM |
|---|---|---|---|---|---|---|---|
| Abalone | Recall | 0.401 | **0.765** | 0.511 | 0.683 | 0.634 | 0.622 |
| | Precision | 0.414 | 0.355 | 0.345 | 0.426 | 0.299 | **0.446** |
| | F1 value | 0.394 | 0.483 | 0.407 | **0.518** | 0.404 | 0.513 |
| | G mean | 0.606 | **0.832** | 0.687 | 0.797 | 0.753 | 0.766 |
| Breast cancer | Recall | 0.862 | 0.940 | 0.902 | 0.9494 | **0.969** | 0.958 |
| | Precision | 0.937 | 0.934 | 0.936 | 0.929 | 0.936 | **0.947** |
| | F1 value | 0.896 | 0.937 | 0.918 | 0.938 | 0.952 | 0.952 |
| | G mean | 0.913 | 0.952 | 0.933 | 0.954 | **0.966** | 0.964 |
| E. coli | Recall | 0.714 | 0.864 | 0.734 | 0.818 | 0.863 | **0.887** |
| | Precision | 0.645 | 0.664 | 0.655 | **0.716** | 0.608 | 0.681 |
| | F1 value | 0.674 | 0.746 | 0.689 | 0.761 | 0.710 | **0.766** |
| | G mean | 0.791 | 0.863 | 0.803 | 0.858 | 0.846 | **0.879** |
| Glass | Recall | 0.817 | 0.863 | 0.790 | 0.871 | 0.859 | **0.880** |
| | Precision | 0.800 | 0.842 | 0.857 | 0.843 | **0.889** | 0.836 |
| | F1 value | 0.800 | 0.849 | 0.817 | 0.850 | **0.867** | 0.852 |
| | G mean | 0.870 | 0.903 | 0.867 | 0.906 | 0.908 | **0.908** |
| Pima | Recall | 0.556 | **0.739** | 0.634 | 0.708 | 0.677 | 0.655 |
| | Precision | 0.604 | 0.596 | 0.551 | 0.603 | 0.568 | **0.626** |
| | F1 value | 0.577 | **0.657** | 0.589 | 0.649 | 0.617 | 0.637 |
| | G mean | 0.667 | **0.730** | 0.677 | 0.726 | 0.700 | 0.716 |
| Vehicle | Recall | 0.898 | 0.949 | 0.935 | 0.962 | 0.967 | **0.969** |
| | Precision | 0.904 | 0.907 | 0.899 | **0.920** | 0.849 | 0.884 |
| | F1 value | 0.900 | 0.926 | 0.917 | **0.940** | 0.903 | 0.924 |
| | G mean | 0.933 | 0.959 | 0.951 | **0.968** | 0.956 | 0.964 |
| Yeast | Recall | 0.674 | 0.806 | 0.782 | **0.809** | 0.775 | 0.758 |
| | Precision | **0.730** | 0.636 | 0.558 | 0.641 | 0.649 | 0.721 |
| | F1 value | 0.700 | 0.710 | 0.650 | 0.714 | 0.706 | **0.736** |
| | G mean | 0.793 | 0.842 | 0.809 | **0.845** | 0.831 | 0.835 |
| Ozone | Recall | 0.035 | 0.360 | 0.358 | 0.362 | **0.499** | 0.280 |
| | Precision | 0.089 | 0.173 | 0.172 | 0.172 | 0.139 | **0.252** |
| | F1 value | 0.047 | 0.228 | 0.228 | 0.228 | 0.214 | **0.256** |
| | G mean | 0.094 | 0.572 | 0.569 | 0.566 | **0.659** | 0.513 |
| Average | Recall | 0.620 | **0.786** | 0.706 | 0.770 | 0.780 | 0.751 |
| | Precision | 0.640 | 0.638 | 0.622 | 0.656 | 0.617 | **0.674** |
| | F1 value | 0.624 | 0.692 | 0.652 | 0.700 | 0.672 | **0.705** |
| | G mean | 0.708 | **0.832** | 0.787 | 0.828 | 0.827 | 0.818 |
| Average Rank | Recall | 1.13 | 4.50 | 2.25 | **4.63** | 4.38 | 4.13 |
| | Precision | 3.50 | 3.38 | 2.63 | 4.00 | 2.50 | **4.75** |
| | F1 value | 1.13 | 4.00 | 2.25 | 4.75 | 3.25 | **5.13** |
| | G mean | 1.13 | **4.63** | 2.13 | **4.63** | 4.00 | 4.38 |
| Average Overall Rank | | 1.72 | 4.13 | 2.31 | 4.50 | 3.53 | **4.59** |

## IV. EEG EXPERIMENTS

To further demonstrate the effectiveness of the ASSOM based method we consider a difficult but practically very useful EEG-based BCI application. The collection of considerable amount of valid EEG data typically has a high cost and often impractical; however, insufficient data have significant impact on the performance defeating use of such methods in real applications.

EEG data helps to assess the states of the brain and hence EEG signals are commonly used in real-world applications [25, 26]. In the EEG-based brain-computer interface (BCI) design, well-recorded data are difficult to collect because most of the subjects are affected by interior and exterior disturbances. These disturbances greatly reduce the quality of the collected data; therefore, the collection of substantial EEG data is always a challenge for constructing BCIs. But use of adequate reliable data are necessary for development of useful systems. The oversampling approach is expected to be an effective method to compensate for insufficient information by generating synthetic samples. In our study, the EEG signals collected from driving task [27].

### A. Driving Task

#### 1) Experiment and subjects

In this study, 33 subjects with normal or corrected vision were recruited for the continuous attention driving experiment. Subjects were asked not to drink alcoholic or caffeinated beverages or participate in strenuous exercise the day before the experiment to ensure that their driving performance could be accurately assessed. Prior to the experiment, all subjects practiced driving in the simulator to be familiar with experimental procedures. This task implemented an immersive driving environment providing a simulated nighttime driving environment on a four-lane highway. Regarding the experimental paradigm, lane-departure events were randomly activated during the simulated driving to cause the car to drift away from the center of the cruising lane (deviation onset). The subjects were instructed to steer the car back (response onset) to the lane center (response offset) as soon as possible after becoming aware of the deviation. The lapse in time between the onset of deviation and response was defined as the reaction time (RT). The level of attention determines the period of RTs, for example 'low fatigue' corresponding to the short RT, and the 'high fatigue' corresponding to the long RT[27].

Table 2. Average SVM performance comparison for different comparative methods

| Dataset | Measure | Original | SVM-*Balanced* | SVM-*light* | SMOTE | ADASYN | MWMOTE | ESPO | ASSOM |
|---|---|---|---|---|---|---|---|---|---|
| Abalone | Recall | 0.202 | **0.769** | 0.143 | 0.567 | 0.422 | 0.447 | 0.655 | 0.477 |
|  | Precision | 0.599 | 0.370 | **0.750** | 0.353 | 0.336 | 0.440 | 0.363 | 0.409 |
|  | F1 value | 0.293 | **0.500** | 0.240 | 0.431 | 0.369 | 0.434 | 0.458 | 0.437 |
|  | G mean | 0.418 | **0.840** | 0.327 | 0.716 | 0.626 | 0.652 | 0.773 | 0.674 |
| Breast cancer | Recall | 0.961 | 0.986 | 0.967 | 0.962 | 0.984 | 0.988 | **0.988** | 0.977 |
|  | Precision | 0.945 | 0.932 | **0.947** | 0.932 | 0.931 | 0.935 | 0.928 | 0.946 |
|  | F1 value | 0.952 | 0.958 | 0.957 | 0.947 | 0.956 | 0.961 | 0.957 | **0.961** |
|  | G mean | 0.965 | 0.958 | 0.957 | 0.962 | 0.972 | **0.975** | 0.973 | 0.973 |
| E. coli | Recall | 0.761 | 0.773 | 0.779 | 0.776 | 0.773 | 0.843 | **0.925** | 0.863 |
|  | Precision | **0.873** | 0.586 | 0.811 | 0.683 | 0.726 | 0.727 | 0.631 | 0.730 |
|  | F1 value | **0.811** | 0.667 | 0.795 | 0.722 | 0.744 | 0.775 | 0.747 | 0.789 |
|  | G mean | 0.857 | 0.673 | 0.795 | 0.829 | 0.837 | 0.869 | 0.878 | **0.882** |
| Glass | Recall | 0.789 | 0.667 | 0.882 | 0.876 | **0.890** | 0.885 | 0.862 | 0.871 |
|  | Precision | 0.840 | 0.800 | **0.918** | 0.850 | 0.835 | 0.874 | 0.843 | 0.873 |
|  | F1 value | 0.799 | 0.727 | **0.900** | 0.857 | 0.855 | 0.875 | 0.850 | 0.870 |
|  | G mean | 0.860 | 0.730 | 0.900 | 0.91 | 0.914 | **0.919** | 0.904 | 0.913 |
| Pima | Recall | 0.536 | 0.685 | 0.571 | 0.613 | 0.560 | 0.643 | **0.735** | 0.694 |
|  | Precision | **0.691** | 0.625 | 0.662 | 0.573 | 0.555 | 0.607 | 0.616 | 0.620 |
|  | F1 value | 0.602 | 0.654 | 0.613 | 0.590 | 0.553 | 0.623 | **0.667** | 0.654 |
|  | G mean | 0.681 | 0.654 | 0.615 | 0.679 | 0.648 | 0.706 | **0.740** | 0.731 |
| Vehicle | Recall | 0.952 | 0.983 | **1.000** | 0.960 | 0.935 | 0.962 | 0.991 | 0.956 |
|  | Precision | 0.939 | 0.862 | 0.765 | 0.934 | 0.941 | 0.906 | 0.872 | **0.951** |
|  | F1 value | 0.946 | 0.918 | 0.867 | 0.946 | 0.935 | 0.931 | 0.927 | **0.953** |
|  | G mean | 0.966 | 0.968 | 0.875 | 0.969 | 0.957 | 0.964 | **0.973** | 0.970 |
| Yeast | Recall | 0.670 | **0.847** | 0.641 | 0.714 | 0.658 | 0.746 | 0.804 | 0.769 |
|  | Precision | **0.810** | 0.537 | 0.809 | 0.678 | 0.591 | 0.646 | 0.624 | 0.670 |
|  | F1 value | **0.733** | 0.658 | 0.716 | 0.695 | 0.621 | 0.690 | 0.702 | 0.715 |
|  | G mean | 0.801 | 0.837 | 0.720 | 0.807 | 0.761 | 0.815 | **0.838** | 0.832 |
| Ozone | Recall | 0.197 | 0.181 | 0.287 | 0.285 | 0.199 | 0.311 | **0.327** | 0.184 |
|  | Precision | 0.250 | 0.217 | 0.186 | 0.260 | 0.283 | **0.288** | 0.167 | 0.276 |
|  | F1 value | 0.209 | 0.212 | 0.223 | 0.265 | 0.225 | **0.293** | 0.219 | 0.214 |
|  | G mean | 0.421 | 0.398 | 0.516 | 0.516 | 0.434 | 0.540 | **0.542** | 0.413 |
| Average | Recall | 0.634 | 0.736 | 0.659 | 0.719 | 0.678 | 0.728 | **0.786** | 0.724 |
|  | Precision | **0.743** | 0.616 | 0.731 | 0.658 | 0.650 | 0.678 | 0.631 | 0.684 |
|  | F1 value | 0.668 | 0.662 | 0.664 | 0.682 | 0.657 | 0.698 | 0.691 | **0.699** |
|  | G mean | 0.746 | 0.757 | 0.713 | 0.799 | 0.769 | 0.805 | **0.828** | 0.799 |
| Average Rank | Recall | 1.88 | 4.75 | 4.13 | 4.25 | 3.38 | 5.75 | **6.88** | 4.75 |
|  | Precision | **6.25** | 2.63 | 6.00 | 3.88 | 3.25 | 5.25 | 2.63 | 6.00 |
|  | F1 value | 4.00 | 3.50 | 4.63 | 3.88 | 3.25 | 5.50 | 4.75 | **6.00** |
|  | G mean | 3.50 | 3.50 | 1.88 | 4.50 | 3.63 | 5.88 | **7.00** | 5.88 |
| Average Overall Rank |  | 3.91 | 3.59 | 4.16 | 4.13 | 3.38 | 5.59 | 5.31 | **5.66** |

*2) EEG signal processing*

During the experiments, the EEG signals were recorded using Ag/AgCl electrodes that were attached to a 32-channel Quik-Cap (Compumedical NeuroScan). Thirty electrodes were arranged according to a modified international 10-20 system, and two reference electrodes were placed on both mastoid bones. The impedance of the electrodes was calibrated under 5kΩ, and the EEG signals recorded at a sampling rate of 500 Hz with 16-bit quantization. Before the data were analyzed, the raw EEG recordings were inspected manually to remove significant artifacts and noisy channels and pre-processed using a digital band-pass filter (1-30 Hz) to remove line noise and artifacts. The EEG signal was estimated using 512-point fast Fourier Transformation (FFT). The step size was set to 1 sec (500 points). Each 512-points sub-window was then transformed to the frequency domain using 512-points FFT, and the mean value of all sub-windows in the frequency domain was calculated as the output of the FFT process. The theta band power of EEG has been identified to distinguish the cognition states: alertness and drowsiness. Furthermore, the EEG coherence, a measure of the degree of similarity of the EEG recorded between pairs of channels, is also considered the patterns to distinguish the states of low fatigue, medium fatigue, and high fatigue.

Table 3 Fatigue state identification of driving task

| EEG power & coherence | State [Accuracy (Mean + SD %)] | $Cluster_{Low\ Fatigue}$ | $Cluster_{Medium\ Fatigue}$ | $Cluster_{High\ Fatigue}$ |
|---|---|---|---|---|
| w/o oversampling | Low Fatigue | **74.8±1.3** | 8.1±1.4 | 16.9±2.8 |
| | Medium Fatigue | 40.7±2.2 | **24.4±5.7** | 34.9±2.9 |
| | High Fatigue | 31.2±2.4 | 23.8±1.0 | **45.0±2.2** |
| w oversampling | Low Fatigue | **78.7±4.3** | 10.2±3.4 | 11.2±2.8 |
| | Medium Fatigue | 37.1±9.2 | **46.4±7.2** | 16.6±4.9 |
| | High Fatigue | 20.7±1.8 | 9.7±2.0 | **69.6±3.2** |
| IR % | Overall | 3.9% | 24% | 24.6% |

*3) Evaluation results*

A multidimensional feature vector consisting of EEG theta power and coherence of 30 channels is used to cluster the accuracy of (low, medium, and high) fatigue states by the Gaussian mixture model (GMM)[29] . In Table 3, we compare the classification performance of grouping into the three fatigue states with and without the oversampling process. In particular, without oversampling, the classifications accuracies of low fatigue, medium fatigue, and high fatigue, are 74.8±1.3%, 24.4±5.7%, and 45.0±2.2%, respectively. However, with the inclusion of oversampling by our ASSOM-based method, the classification performance on the low fatigue, medium fatigue, and high fatigue classes improved to 75.8±1.4%, 32.7±2.4%, and 55.9±1.0%, respectively.

V. DISCUSSION AND CONCLUSION

In this paper, we have elaborated the ASSOM based minority oversampling method that was proposed in [20]. Since our intention is to generate synthetic samples that represent the original data better, we have used a different discounting kernel function as the distance of a non-winner module and winner module changes on the ASSOM array. Since ASSOM finds subspaces invariant to rotation and translation, exploiting of those subspaces to generate synthetic samples is found to be quite effective for conventionally used benchmark datasets.

Considering the wide range of imbalance situations in the neuroscientific and neurological data, limited studies investigated the imbalance learning problem in the context of analyzing EEG signals and BCI. A recent study showed the possibility to generate artificial EEG signals with a generative adversarial network (GAN)[28]. However, we did not exploit the notion of GAN to generate time-series samples. The ASSOM finds subspaces that capturer some invariant properties of the data which are found to be effective to generate synthetic data. But we note here that our method does not generate raw EEG signal, but features extracted from EEG.

To summarize, this work extended the study on generation of synthetic samples to deal with the imbalance classification problem using an ASSOM based algorithm reported in [20]. The study in [20] used Kohonen's ASSOM[15,16] for synthetic sample generation. This was not clear in due to lack of appropriate referencing. We have clarified that issue here to avoid confusion of readers. The sample generation algorithm was not clearly described in [20]. Here we have provided a detailed algorithm so that anyone can implement it. The kernel function used in the objective function to reduce the importance of the non-winners in the ASSOM layer generally is a decreasing function of the distance between the winner and a non-winner on the ASSOM layer, which is commonly done also in case of SOM. But here, we used a different kernel function keeping our goal of synthetic sample generation in mind. We have compared the performance of the ASSOM-based method on several benchmark data sets using two popular classifiers and several performance indices. The performance of our approach is found to be better. For many EEG applications, obtaining many samples from the positive class is difficult. Often the captured data are noisy. To deal with such problems we have applied our method on an EEG-based BCI application: Driver's fatigue states classification. Our results demonstrate the effectiveness of the ASSOM-based minority over-sampling method.

ACKNOWLEDGMENT

The authors would like to thank the Ministry of Science and Technology, Taiwan, for financially supported this work under contract no. MOST 106-2218-E-009-027-MY3 and MOST 108-2221-E-009 -120 -MY2.

REFERENCES

[1] V. Chandola, A. Banerjee, and V. Kumar, "Anomaly detection: A survey," *ACM computing surveys (CSUR),* vol. 41, no. 3, pp. 1-58, 2009.
[2] S. C. Tan, J. Watada, Z. Ibrahim, and M. Khalid, "Evolutionary Fuzzy ARTMAP Neural Networks for Classification of

Semiconductor Defects," (in English), *Ieee Transactions on Neural Networks and Learning Systems,* vol. 26, no. 5, pp. 933-950, May 2015.

[3] Q. Kang, X. S. Chen, S. S. Li, and M. C. Zhou, "A Noise-Filtered Under-Sampling Scheme for Imbalanced Classification," (in English), *Ieee Transactions on Cybernetics,* vol. 47, no. 12, pp. 4263-4274, Dec 2017.

[4] P. Lim, C. K. Goh, and K. C. Tan, "Evolutionary Cluster-Based Synthetic Oversampling Ensemble (ECO-Ensemble) for Imbalance Learning," (in English), *Ieee Transactions on Cybernetics,* vol. 47, no. 9, pp. 2850-2861, Sep 2017.

[5] H. He and E. A. Garcia, "Learning from imbalanced data," *IEEE Transactions on knowledge and data engineering,* vol. 21, no. 9, pp. 1263-1284, 2009.

[6] R. Batuwita and V. Palade, "FSVM-CIL: Fuzzy Support Vector Machines for Class Imbalance Learning," (in English), *Ieee Transactions on Fuzzy Systems,* vol. 18, no. 3, pp. 558-571, Jun 2010.

[7] Z. H. Zhou and X. Y. Liu, "Training cost-sensitive neural networks with methods addressing the class imbalance problem," (in English), *Ieee Transactions on Knowledge and Data Engineering,* vol. 18, no. 1, pp. 63-77, Jan 2006.

[8] S. J. Yen and Y. S. Lee, "Cluster-based under-sampling approaches for imbalanced data distributions,", *Expert Systems with Applications,* vol. 36, no. 3, pp. 5718-5727, Apr 2009.

[9] N. V. Chawla, K. W. Bowyer, L. O. Hall, and W. P. Kegelmeyer, "SMOTE: Synthetic minority over-sampling technique," (in English), *Journal of Artificial Intelligence Research,* vol. 16, pp. 321-357, 2002.

[10] H. Han, W. Y. Wang, and B. H. Mao, "Borderline-SMOTE: A new over-sampling method in imbalanced data sets learning," (in English), *Advances in Intelligent Computing, Pt 1, Proceedings,* vol. 3644, pp. 878-887, 2005.

[11] N. V. Chawla, A. Lazarevic, L. O. Hall, and K. W. Bowyer, "SMOTEBoost: Improving prediction of the minority class in boosting," (in English), *Knowledge Discovery in Databases: Pkdd 2003, Proceedings,* vol. 2838, pp. 107-119, 2003.

[12] S. Barua, M. M. Islam, X. Yao, and K. Murase, "MWMOTE-Majority Weighted Minority Oversampling Technique for Imbalanced Data Set Learning," (in English), *Ieee Transactions on Knowledge and Data Engineering,* vol. 26, no. 2, pp. 405-425, Feb 2014.

[13] H. B. He, Y. Bai, E. A. Garcia, and S. T. Li, "ADASYN: Adaptive Synthetic Sampling Approach for Imbalanced Learning," (in English), *2008 Ieee International Joint Conference on Neural Networks, Vols 1-8,* pp. 1322-1328, 2008.

[14] H. Cao, X. L. Li, D. Y. K. Woon, and S. K. Ng, "Integrated Oversampling for Imbalanced Time Series Classification," (in English), *Ieee Transactions on Knowledge and Data Engineering,* vol. 25, no. 12, pp. 2809-2822, Dec 2013.

[15] T. Kohonen, "Emergence of invariant-feature detectors in the adaptive-subspace self-organizing map," (in English), *Biological Cybernetics,* vol. 75, no. 4, pp. 281-291, Oct 1996.

[16] T. Kohonen, S. Kaski, H. Lappalainen, and J. Saljärvi, "The adaptive-subspace self-organizing map (ASSOM)," in *International Workshop on Self—Organizing Maps (WSOM'97), Helsinki*, 1997, pp. 191-196

[17] H. Kawano, K. Horio, and T. Yamakawa, "A pattern classification method using kernel adaptive-subspace self-organizing map," *IEEJ Transactions on Electronics, Information and Systems,* vol. 125, no. 1, pp. 149-150, 2005.

[18] H. Kawano, T. Yamakawa, and K. Horio, "Kernel-based adaptive-subspace self-organizing map as a nonlinear subspace pattern recognition," in *Proceedings World Automation Congress, 2004.*, 2004, vol. 18, pp. 267-272: IEEE.

[19] C.-T. Lin, Hsieh, T.Y., Liu, Y.T., Lin, Y.Y., Fang, C.N., Wang, Y.K., Yen, G., Pal, N.R. and Chuang, C.H, "Minority Oversampling in Kernel Adaptive Subspaces for Class Imbalanced Datasets," *IEEE Transactions on Knowledge & Data Engineering, 30*(5), 950-962, 2017

[20] Y.-T. Liu, N. R. Pal, S.-L. Wu, T.-Y. Hsieh, and C.-T. Lin, "Adaptive subspace sampling for class imbalance processing," in *2016 International Conference on Fuzzy Theory and Its Applications (iFuzzy)*, 2016, pp. 1-5: IEEE.

[21] A. K. Jain, J. Mao, and K. M. Mohiuddin, "Artificial neural networks: A tutorial," *Computer,* vol. 29, no. 3, pp. 31-44, 1996.

[22] Y. Liu and Y. F. Zheng, "Soft SVM and its application in video-object extraction," *IEEE Transactions on Signal Processing,* vol. 55, no. 7, pp. 3272-3282, 2007.

[23] A. Asuncion and D. Newman, "UCI machine learning repository," ed, 2007.

[24] J. Alcalá-Fdez *et al.*, "Keel data-mining software tool: data set repository, integration of algorithms and experimental analysis framework," *Journal of Multiple-Valued Logic & Soft Computing,* vol. 17, 2011.

[25] L. Z. Bi, H. K. Wang, T. Teng, and C. T. Guan, "A Novel Method of Emergency Situation Detection for a Brain-Controlled Vehicle by Combining EEG Signals With Surrounding Information," (in English), *Ieee Transactions on Neural Systems and Rehabilitation Engineering,* vol. 26, no. 10, pp. 1926-1934, Oct 2018.

[26] T. Chouhan, N. Robinson, A. P. Vinod, K. K. Ang, and C. T. Guan, "Wavlet phase-locking based binary classification of hand movement directions from EEG," (in English), *Journal of Neural Engineering,* vol. 15, no. 6, Dec 2018.

[27] K. C. Huang, C. H. Chuang, Y. K. Wang, C. Y. Hsieh, J. T. King, and C. T. Lin, "The effects of different fatigue levels on brain-behavior relationships in driving," (in English), *Brain and Behavior,* vol. 9, no. 12, Dec 2019.

[28] K. G. Hartmann, R. T. Schirrmeister, and T. Ball, "EEG-GAN: Generative adversarial networks for electroencephalograhic (EEG) brain signals," *arXiv preprint arXiv:1806.01875,* 2018.

[29] C.-H. Chuang, Z. Cao, P.-T. Chen, C.-S. Huang, N. R. Pal, and C.-T. Lin, "Dynamically weighted ensemble-based prediction system for adaptively modeling driver reaction time," *arXiv preprint arXiv:1809.06675,* 2018.

[30] Kohonen T, Kaski S, Lappalainen H. Self-organized formation of various invariant-feature filters in the adaptive-subspace SOM. Neural computation. 1997 Aug 15;9(6):1321-44.